\documentclass[sigconf,natbib=true]{acmart}
\usepackage{algorithm}
\usepackage{algorithmic}
\usepackage{dsfont}
\usepackage{makecell}
\usepackage{enumitem}
\usepackage{multirow}
\usepackage{booktabs}
\usepackage{geometry}
\usepackage{subcaption}
\usepackage{amsmath}
\usepackage{colortbl}
\usepackage{tcolorbox}
\AtBeginDocument{%
  }

\begin{document}

\title{Dual-Tree LLM-Enhanced Negative Sampling for Implicit Collaborative Filtering}


\begin{abstract}
 Negative sampling is a pivotal technique in implicit collaborative filtering (CF) recommendation, enabling efficient and effective training by contrasting observed interactions with sampled unobserved ones.
 Recently, large language models (LLMs) have shown promise in recommender systems; however, research on LLM-empowered negative sampling remains underexplored. 
  Existing methods heavily rely on textual information and task-specific fine-tuning, limiting practical applicability. 
  To this end, we propose a text-free and fine-tuning-free \textbf{D}ual-\textbf{T}ree \textbf{L}LM-enhanced \textbf{N}egative \textbf{S}ampling method (DTL-NS).
  It consists of two modules: (i) an offline false negative identification module that leverages hierarchical index trees to transform collaborative structural and latent semantic information into structured item-ID encodings for LLM inference, enabling accurate identification of false negatives; and (ii) a multi-view hard negative sampling module that combines user–item preference scores with item–item hierarchical similarities from these encodings to mine high-quality negatives, thus improving models’ discriminative ability.
  Extensive experiments demonstrate the effectiveness of DTL-NS. 
  Moreover, DTL-NS shows broad applicability across different implicit CF models, negative sampling methods, and LLMs, consistently enhancing recommendation performance.
  
\end{abstract}



\author{Jiayi Wu}
\affiliation{
	\institution{Beijing Institute of Technology}
	\city{Beijing}
	\country{China}
}

\author{Zhengyu Wu}
\affiliation{
	\institution{Beijing Institute of Technology}
	\city{Beijing}
	\country{China}
}

\author{Xunkai Li}
\affiliation{
	\institution{Beijing Institute of Technology}
	\city{Beijing}
	\country{China}
}

\author{Rong-Hua Li}
\email{lironghuabit@126.com}
\affiliation{
	\institution{Beijing Institute of Technology}
	\city{Beijing}
	\country{China}
}
\authornote{Corresponding author.}

\author{Guoren Wang}
\affiliation{
	\institution{Beijing Institute of Technology}
	\city{Beijing}
	\country{China}
}


\maketitle

\section{Introduction}

Recommender systems are widely used in online platforms such as e-commerce~\cite{zhao2022e-commerce} and news services~\cite{wu2023news} to deliver personalized content and improve user experience~\cite{zhang2025personalized}. 
Since explicit feedback (e.g., ratings and reviews) is often sparse and difficult to obtain~\cite{cai2022ratings}, practical systems typically rely on implicit feedback derived from user behavior logs such as clicks and views~\cite{xin2023click}. 
Thus, implicit collaborative filtering (CF) has become a central modeling paradigm in modern recommender systems~\cite{yin2019collaborative}. 
However, implicit feedback contains a large amount of unobserved user–item interactions~\cite{johnson2014implicitdata}, and treating all unobserved items as negatives is computationally inefficient~\cite{Chen2023nonsampling}. 
Negative sampling is therefore widely adopted to select a small subset of unobserved items as training negatives, reducing computational cost while providing effective supervision~\cite{yang2024nssurvey}.

Existing negative sampling methods~\cite{DNS, MixGCF, ANS} mainly focus on mining hard negatives to improve the model’s discrimination between positives and negatives. 
However, hard negatives are often highly similar to positive items, and some may actually reflect users’ latent preferences that remain unobserved due to limited exposure, thus introducing noisy supervision in the form of false negatives. 
Prior work~\cite{GDNS, SRNS} mitigates this issue using heuristic rules derived from training-time observations, such as loss variance, but these heuristics are often dataset-dependent and mainly aim to reduce the likelihood of sampling false negatives, while overlooking the potential of false negatives to serve as latent positives. 
Therefore, effectively identifying potential false negatives from unobserved data is crucial for improving negative sampling quality.

Recently, large language models (LLMs) have been increasingly adopted in recommender systems due to their strong language understanding, instruction-following, and semantic discrimination capabilities.
Prior work largely falls into two lines based on the information sources leveraged by LLMs: (1) text-based approaches~\cite{wei2024llmrec, ren2024rlmrec} that use LLMs as semantic encoders or profile generators to enrich user/item representations and fuse or align them with CF-based ID embeddings, but they heavily depend on textual metadata; and (2) ID-only approaches~\cite{li2024e4srec, chen2025llm4idrec, zhang2024bin4llm} that treat item IDs as tokens and use LLMs as backbone recommenders to learn users’ sequential preferences via instruction-based fine-tuning, but they incur considerable training costs.
Although LLMs have improved recommendation performance, their use in negative sampling remains underexplored. 
Among limited studies, HNLMRec~\cite{zhao2025HNLMRec} injects user/item profiles and CF signals into prompts, and fine-tunes an LLM to generate negative embeddings. 
However, this design still suffers from two limitations: (1) \textbf{text dependency}, as it relies on textual metadata (e.g., titles/descriptions) for prompts or semantic constraints; and (2) \textbf{fine-tuning dependency}, since supervised fine-tuning (even with LoRA) still requires forward passes through the entire LLM and backpropagation to optimize the adapter parameters, incurring non-trivial training cost. 
Moreover, HNLMRec reports that without fine-tuning, its LLM-based negative sampler fails to outperform state-of-the-art samplers.

\begin{figure}[t]
	\includegraphics[width=\linewidth]{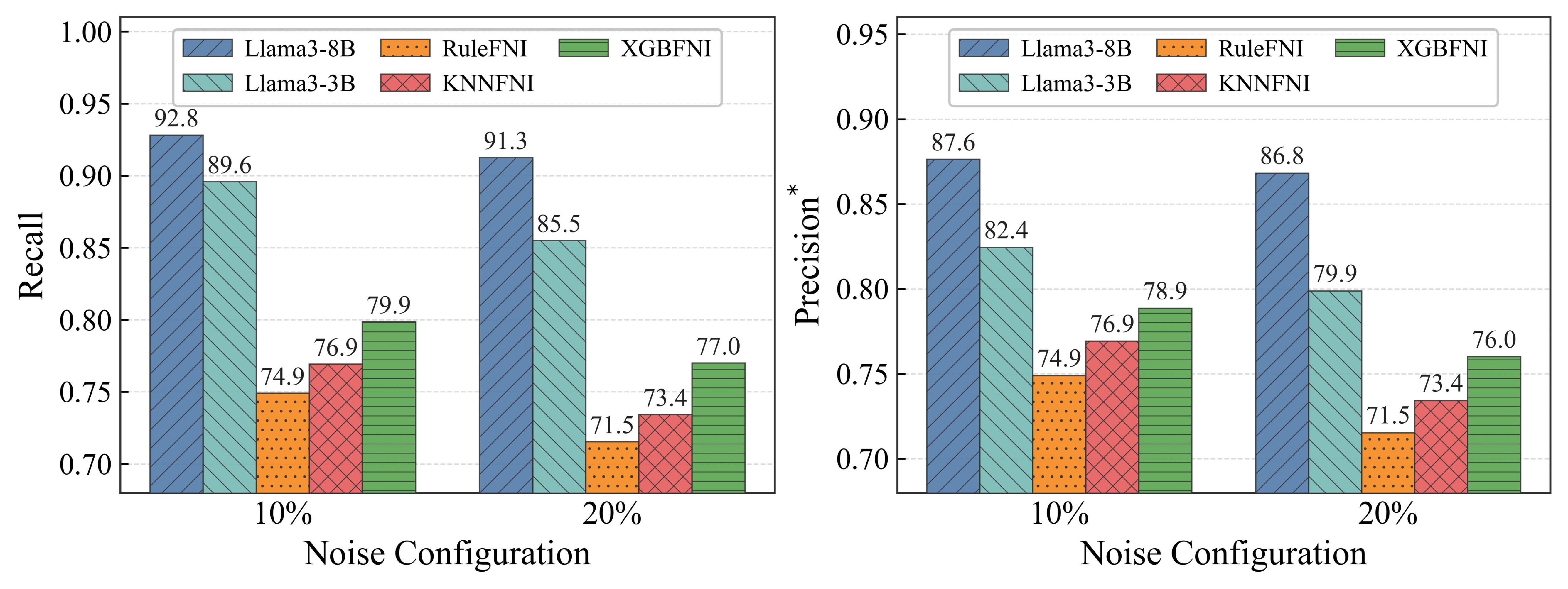}
	
	\vspace{-0.3cm}
	
\caption{False-negative identification performance comparison on Amazon-sports.}

	\vspace{-0.5cm}

\end{figure}

To address the above limitations, we propose DTL-NS, a text-free and fine-tuning-free LLM-enhanced negative sampling framework. 
DTL-NS constructs dual hierarchical index trees from collaborative structural and latent semantic information, and converts each item into a compact and interpretable tree-path encoding. 
Based on these structured encodings, the LLM performs offline classification over unobserved candidates to identify potential false negatives. 
To examine its effectiveness, we conduct a controlled diagnostic study by randomly removing 10\% and 20\% of training positives as known false negatives, mixing them with an equal number of randomly sampled unobserved items as background candidates, and asking the LLM to classify each candidate using only its tree-path encodings.
In addition to LLM-based classification, we compare three no-LLM variants under the same controlled setting: RuleFNI, KNNFNI, and XGBFNI, corresponding to rule-based, retrieval-based, and supervised machine-learning-based methods, respectively. Details are provided in Section~4.4. 
For fair comparison, we adopt a count-matched ranking protocol, where each no-LLM variant ranks candidates by its false-negative score or predicted positive probability and predicts the top-$K_u$ candidates as positives, with $K_u$ denoting the number of simulated false negatives for user $u$. 
Under this protocol, the predicted-positive count matches the number of known false negatives for each user, so Recall and Precision$^{*}$ are numerically identical for these variants.
Figure~1 reports false-negative recall (Recall) and conservative precision lower bound (Precision$^{*}$), where Precision$^{*}$ is conservative because the randomly sampled unobserved items may also contain false negatives. 
The results show that LLM-based classifiers substantially outperform the no-LLM variants.
In particular, Llama3-8B~\cite{grattafiori2024llama} consistently achieves Recall above 90\% and Precision$^{*}$ above 85\%, and also outperforms Llama3-3B, suggesting that stronger LLMs provide more reliable false-negative identification. 
These results demonstrate the advantage of LLM-based false-negative identification, showing that LLMs can effectively identify a substantial portion of simulated false negatives without relying on textual metadata or fine-tuning.
DTL-NS then treats the LLM-identified false negatives as positives to augment training, enabling the model to better exploit latent preference signals in unobserved data. 
During training, DTL-NS further integrates dual-tree item-item hierarchical similarity with user-item preference signals to perform multi-view hard negative sampling, thereby yielding higher-quality hard negatives.

Our contributions are summarized as follows:
\setlist[itemize,1]{left=0pt}
\begin{itemize}

\item We design a novel LLM-based false-negative identification module that leverages hierarchical index trees to transform collaborative structural and latent semantic information into structured ID representations, enabling accurate identification of false negatives without relying on textual metadata or fine-tuning.

\item We develop a multi-view hard negative sampling module that combines user–item preference scores with item–item hierarchical similarity to select negatives structurally and semantically closer to positives, improving the model’s discriminative ability.

\item By integrating the above two modules, we propose DTL-NS. Extensive experiments show that DTL-NS consistently outperforms strong baselines and further improves various implicit CF models and negative sampling methods in a plug-and-play manner.

\end{itemize}

\section{Related Work}

\subsection{Negative Sampling in Recommendation}

Existing negative sampling methods mainly aim to mine harder negatives to improve the model’s discrimination between positives and negatives. They can be roughly divided into two categories. 
The first samples hard negatives from real unobserved items, such as DNS~\cite{DNS} and DNS(M,N)~\cite{DNS(MN)}, which select high-scoring items according to model-predicted user-item preference scores. 
The second synthesizes hard negatives in the embedding space, such as MixGCF~\cite{MixGCF} and ANS~\cite{ANS}, which generate negative embeddings via mixup~\cite{mixup} or disentangling~\cite{DENS} operations and then select hard negatives using score-based strategies. 
However, these methods mainly rely on the single-view signal of user--item preference scores, while overlooking item-item collaborative and semantic relations. 
As a result, they struggle to select negatives that are structurally or semantically close to the positives. 
In contrast, DTL-NS explicitly captures item–item collaborative and semantic relations via dual-tree path encoding similarities and integrates them with user–item preference signals to construct a multi-view criterion for hard negative sampling. This multi-view design enables the sampler to discover harder-to-distinguish negatives and thereby further enhance the model’s discriminative capability.

Beyond hard negative sampling, another line of work focuses on false-negative risk, since unobserved interactions may not be true negatives and treating them all as negatives can introduce noisy supervision. 
Representative methods adopt different strategies to mitigate this risk: SRNS~\cite{SRNS} and GDNS~\cite{GDNS} exploit training dynamics such as losses or gradients; BNS~\cite{liu2023bayesian} estimates true-negative probabilities from item-popularity priors and model-dependent rankings; RecNS~\cite{RecNS} and RealHNS~\cite{RealHNS} leverage external signals such as exposed-but-unclicked data or cross-domain data; and AHNS~\cite{AHNS} adaptively adjusts negative hardness. 
However, these methods mainly aim to avoid sampling false negatives, while overlooking their potential as latent positives. 
In contrast, DTL-NS performs offline LLM-based false-negative identification over structured item encodings and treats the identified false negatives as positives, thereby strengthening positive supervision and improving the accuracy of preference learning.

\begin{figure*}[tbp]
	\centering

	\includegraphics[width=\linewidth]{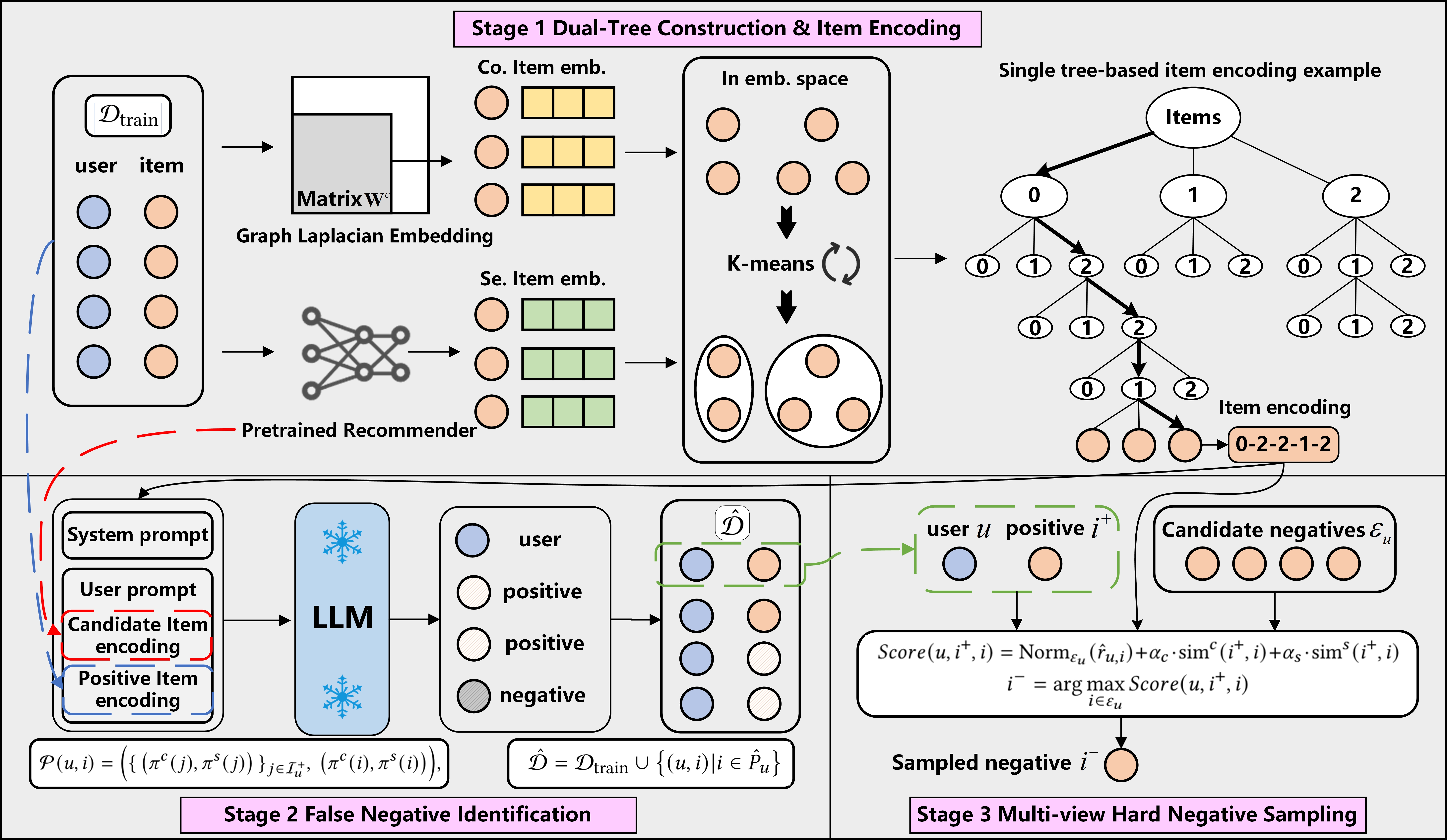}
	
		\vspace{-0.1cm}
	
\caption{The DTL-NS framework. We depict a single tree-based item encoding example, since the dual trees use node embeddings from different sources, yet share the same subsequent tree construction and item encoding procedures.}

\vspace{-0.2cm}

\end{figure*}

\subsection{LLM-based Recommendation and Negative Sampling}

LLM-based recommendation methods have primarily focused on text-driven LLM recommenders. Existing work can be grouped into two main categories:
(1) LLM as an enhancer, where LLMs leverage textual metadata such as titles and reviews to generate user/item profiles or extract semantic representations, which are then jointly optimized with interaction signals to enhance recommendation performance (e.g., LLM4Rec~\cite{wei2024llmrec}, RLM4Rec~\cite{ren2024rlmrec}, AlphaRec~\cite{AlphaRec}); and (2) LLM as a recommender, where the recommendation task is reformulated into an instruction-driven prediction task, and the LLM is fine-tuned on prompt–response pairs constructed from textual information and downstream objectives, enabling instruction-based recommendation (e.g., P5~\cite{P5}, TALLRec~\cite{TALLRec}, InstructRec~\cite{InstructRec}).
Beyond the above text-driven approaches, another important line of work explores ID-driven LLM recommenders, which inject item IDs and collaborative signals into LLMs, either as discrete tokens or continuous embeddings, and fine-tune LLMs to learn users’ interaction patterns, thereby improving recommendation performance (e.g., BinLLM~\cite{zhang2024bin4llm}, LLM4IDRec~\cite{chen2025llm4idrec}, E4SRec~\cite{li2024e4srec}).
Overall, the above methods either rely on high-quality textual metadata to introduce semantic knowledge, or require task-specific fine-tuning of LLMs, which incurs additional data construction and training costs. 
More importantly, they primarily focus on how to use LLMs for representation learning and prediction in recommendation, whereas how to leverage LLMs to empower negative sampling, a key training component of recommender, remains largely unexplored.


In the limited work on LLM-empowered negative sampling, HNLMRec~\cite{zhao2025HNLMRec} injects user/item textual profiles and pretrained CF embeddings~\cite{lightgcn} into prompts, and fine-tunes an LLM with MixGCF-generated hard negative embeddings to generate negatives for downstream recommendation. 
However, it still depends on textual metadata and task-specific fine-tuning. 
In contrast, DTL-NS performs offline false-negative identification via text-free, fine-tuning-free LLM-based classification, thereby alleviating false-negative risk at the source and substantially improving its applicability across datasets with limited textual metadata, while also reducing the computational burden by eliminating LLM fine-tuning.

\section{Methodology}

In this section, we first formalize the problem and then present DTL-NS. As shown in Figure~2, DTL-NS consists of three stages: (1) dual-tree construction \& item encoding, which provides foundational structural/semantic encodings; (2) false-negative identification, which, together with the encodings produced in Stage 1, forms our first module, Dual-Tree LLM-based False Negative Identification; and (3) multi-view hard negative sampling, which, combined with the same encodings, constitutes our second module, Dual-Tree Guided Multi-View Hard Negative Sampling.

\vspace{-0.1cm}

\subsection{Problem Formalization}

In implicit CF, let $\mathcal{U}$, $\mathcal{I}$, and $\mathcal{D} \! \subseteq \! \mathcal{U} \! \times \! \mathcal{I}$ denote the user set, the item set, and the set of observed interactions.
The goal is to learn a scoring function $S_\theta(u,i)$ that ranks observed items above unobserved ones for each user. 
Due to the lack of explicit negative feedback, negative sampling is adopted: for each positive pair $(u,i^+) \in \mathcal{D}$, a negative $i^-$ is sampled from $q(\cdot \mid u,i^+)$, and the model is optimized with the Bayesian Personalized Ranking~\cite{BPR} (BPR) loss:
\begin{equation}
\ell  =  - \sum\limits_{\left( {u,{i^ + }} \right) \in \mathcal{D},{i^ - } \sim q\left( { \cdot |u,{i^ + }} \right)} {\log \sigma \left( {{S_\theta }(u,{i^ + }) - {S_\theta }(u,{i^ - })} \right)}.
\end{equation}
However, some unobserved items are not true negatives, and sampling them as negatives may introduce noisy supervision. To mitigate this issue, we introduce DTL-FNI, which we formalize as a function $\Phi(\cdot)$. Given a user $u$ and the unobserved item set $\mathcal{I} \setminus \mathcal{I}_u^+$, $\Phi$ outputs a set of detected false negatives, which are then converted into new positive feedback to augment the training set $\mathcal{D}_{\text{train}}$.
\begin{equation}
{\hat P_u} = \Phi \left( {u,I\backslash I_u^ + } \right) ,\quad \quad  \hat{\mathcal{D}} = \mathcal{D}_{\text{train}} \cup \left\{ {(u,i)|i \in {{\hat P}_u}} \right\}
\end{equation}
where $\mathcal{I}_u^+$ denotes the items observed by $u$. Finally, we introduce DT-MHNS to redefine the negative sampling distribution $q\left( { \cdot |u,{i^ + }} \right)$ in a multi-view manner to mine harder negatives.

\subsection{Dual-Tree LLM-based False Negative Identification (DTL-FNI)}

To avoid the high cost of repeatedly invoking LLM inference during training, DTL-FNI runs entirely offline: it identifies false negatives once before training, treats them as positives, and thus mitigates false-negative risk at its source. 
Furthermore, to enhance applicability, we target a text-free and fine-tuning-free setting. 
However, directly feeding raw item IDs into an LLM is ineffective, as IDs are merely discrete symbols without interpretable semantics~\cite{hua2023index, chen2025llm4idrec}. 
Serializing item embeddings into numerical tokens is also problematic: continuous high-dimensional embeddings are misaligned with the LLM’s discrete token space and serialization incurs prohibitive token overhead~\cite{zhao2025HNLMRec}. 
Continuous prompt injection, where item embeddings are injected into the LLM’s hidden space via projection, also requires additional adapter modules and still lacks interpretable semantic anchors~\cite{li2024e4srec, xu2025slmrec}. 
These limitations make it difficult for the LLM to perform effective inference without fine-tuning.

To this end, DTL-FNI constructs hierarchical index trees and encodes items as compact, interpretable path encodings, enabling efficient and effective offline LLM-based classification.
Specifically, we build a collaborative-structure tree and a latent semantic tree: the former captures item–item structural proximity induced by user interactions, while the latter captures higher-order semantic similarity in the recommender embedding space. 
These two trees provide complementary views, allowing the path encodings jointly reflect interaction co-occurrence and latent relational patterns, offering richer relational cues for false-negative identification.

\textbf{Collaborative-structure tree construction and encoding.}
We first construct an item similarity matrix based on interactions in $\mathcal{D}_{\text{train}}$. Since measuring similarity solely by co-occurrence counts introduces popularity bias, we adopt the Jaccard coefficient~\cite{niwattanakul2013using} to normalize the overlap between user sets, enabling a more robust estimation of item similarity. 
For any two items $i, j \in \mathcal{I}$ with their interacting user sets $U(i)$ and $U(j)$, their similarity is computed as:
\begin{equation}
W_{ij}^{c} = \frac{{|U\left( i \right) \cap U\left( j \right)|}}{{|U\left( i \right) \cup U\left( j \right)|}},\;\;i \ne j.
\end{equation}
This yields the item-item similarity matrix $\mathbf{W}^{c} \in \mathbb{R}^{|\mathcal{I}| \times |\mathcal{I}|}$. Since $\mathbf{W}^{c}$ is highly sparse and local edge weights may be corrupted by noisy interactions (e.g., incidental clicks or accidental touches), we adopt graph Laplacian spectral embedding~\cite{ma2023laplacian} to learn a low-dimensional, structure-preserving representation from a global perspective, which also facilitates subsequent clustering and hierarchical tree construction. 
Specifically, we construct the normalized graph Laplacian matrix:
\begin{equation}
	\mathbf{L}^{c} = \mathbf{I} - \mathbf{D}^{-\frac{1}{2}} \mathbf{W}^{c} \mathbf{D}^{-\frac{1}{2}}
\end{equation}
where $\mathbf{D}$ is the degree matrix. We then take the $d$ smallest non-trivial eigenvectors of $\mathbf{L}^{c}$ and stack them column-wise to obtain the item embedding matrix $\mathbf{E}^{c} \in \mathbb{R}^{|\mathcal{I}| \times d_c}$. To compress continuous item embeddings into a discrete hierarchical structure for encoding, we build a $k$-ary hierarchical index tree $\mathcal{T}^{c}$ on top of $\mathbf{E}^{c}$. Specifically, at each node we run $k$-means~\cite{ahmed2020k} on the embeddings of the items assigned to that node, partitioning them into $k$ child clusters, and recursively apply this procedure to each child cluster until the number of items in a node does not exceed $m$. In the resulting tree $\mathcal{T}^{c}$, each item $i$ is associated with a root-to-leaf path:
\begin{equation}
	\pi^{c}(i) = \bigl(n^{c}_1, n^{c}_2, \ldots, n^{c}_{L_c}\bigr),
\end{equation}
where $n^{c}_\ell$ denotes the cluster node index at level $\ell$ and $L_c$ is the depth of $\mathcal{T}^{c}$. $\pi^{c}(i)$ serves as the collaborative-structure encoding of item $i$, which is used for subsequent offline LLM classification.

\textbf{Latent semantic tree construction and encoding.}
Since the collaborative-structure tree mainly captures user co-occurrence patterns and local neighborhood relations but struggles to model higher-order semantic associations, we construct a latent semantic tree $\mathcal{T}^{s}$ to encode higher-order semantic relations in the recommender’s embedding space as a complementary view. Specifically, we first pre-train an implicit CF model (e.g., LightGCN~\cite{lightgcn}) to obtain a semantic embedding for each item. We then stack all item embeddings into a matrix $\mathbf{E}^{s} \in \mathbb{R}^{|\mathcal{I}| \times d_s}$ and construct the latent semantic tree $\mathcal{T}^{s}$ using the same $k$-ary recursive clustering procedure as in the collaborative-structure tree. In the resulting tree $\mathcal{T}^{s}$, each item $i$ is associated with a root-to-leaf path:
\begin{equation}
	\pi^{s}(i) = \bigl(n^{s}_1, n^{s}_2, \ldots, n^{s}_{L_s}\bigr)
\end{equation}
which serves as the latent semantic encoding of item $i$.

Algorithm~\ref{alg:offline_dualtree} describes the dual-tree construction and item encoding procedure. (Lines 1–6) derive structural embeddings via interaction-based item similarities and Laplacian spectral embedding, then build the collaborative structural tree and encode items accordingly; (Lines 7–11) obtain semantic item embeddings from a pretrained recommender, then construct the latent semantic tree and derive the corresponding item encodings.

\begin{algorithm}[t]
	
	\renewcommand{\algorithmicrequire}{\textbf{Input:}}
	\renewcommand{\algorithmicensure}{\textbf{Output:}}
	\caption{Dual-Tree Construction \& Item Encoding}
	\label{alg:offline_dualtree}
	\begin{algorithmic}[1]
		\REQUIRE Training set $\mathcal{D}_{\text{train}}\subseteq \mathcal{U}\times\mathcal{I}$; pretrained recommender $g$; 
		tree branching factor $k$; leaf size threshold $m$; spectral embedding dimension $d$.
		\ENSURE Collaborative path codes $\pi^c(\cdot)$, semantic path codes $\pi^s(\cdot)$.
		
		\STATE Construct the item-item collaborative similarity matrix $W^c$ from $\mathcal{D}_{\text{train}}$ using (3).
		\STATE Compute structure embeddings $E^c$ via graph laplacian spectral embedding (keep the smallest $d$ non-trivial eigenvectors).
		\STATE $T^c \leftarrow \mathrm{BuildKaryTree}(E^c, k, m)$.
		
		\FOR{each item $i \in \mathcal{I}$}
		\STATE Extract root-to-leaf cluster-index sequence of $i$ in $T^c$ as $\pi^c(i)$.
		\ENDFOR
		
		\STATE Obtain semantic item embeddings $E^s$ from recommender $g$.
		\STATE $T^s \leftarrow \mathrm{BuildKaryTree}(E^s, k, m)$.
		
		\FOR{each item $i \in \mathcal{I}$}
		\STATE Extract root-to-leaf cluster-index sequence of $i$ in $T^s$ as $\pi^s(i)$.
		\ENDFOR
		
		\STATE {\bf Notes:} $\mathrm{BuildKaryTree}(\cdot)$ recursively partitions embeddings via $k$-means until leaf node size $\le m$.
		
	\end{algorithmic}
\end{algorithm}

\textbf{LLM classification.}
We feed the collaborative structure and latent semantic path encodings of unobserved items into the LLM, which performs offline binary classification for false negative identification. 
Since the number of unobserved items is typically enormous, directly applying the LLM to classify all candidates is infeasible. Following prior work, we use a pretrained recommender to construct a finite candidate set $\mathcal{C}_u$ for each user $u$. Items in $\mathcal{C}_u$ have high predicted scores and thus are likely to include both false negatives and hard negatives, providing a more effective candidate space for subsequent offline LLM classification.
For each candidate item $i \in \mathcal{C}_u$, we construct an input prompt:
\begin{equation}
	\mathcal{P}(u,i) = \Bigl(\{\,\bigl(\pi^{c}(j), \pi^{s}(j)\bigr)\,\}_{j \in \mathcal{I}_u^+},\ \bigl(\pi^{c}(i), \pi^{s}(i)\bigr)\Bigr),
\end{equation}
which consists of the dual-tree path encodings of the user’s historical items and the candidate item $i$. We then feed this prompt together with a system prompt $\mathcal{S}_P$ into the LLM to decide whether $i$ is a false negative for user $u$:
\begin{equation}
	\hat{y}_{u,i} = f_{\mathrm{LLM}}\bigl(\mathcal{P}(u,i), \mathcal{S}_P\bigr), \quad
	\hat{y}_{u,i} \in \{\texttt{positive}, \texttt{negative}\}.
\end{equation}
Here, the \texttt{positive} label corresponds to the false negative class. We collect the candidates predicted as \texttt{positive} into the detected false negative set:
\begin{equation}
	\hat{P}_u = \Phi(u, \mathcal{C}_u) = \{\, i \in \mathcal{C}_u \mid \hat{y}_{u,i} = \texttt{positive} \,\}.
\end{equation}
Finally, we add $\hat{P}_u$ to the training set for subsequent model training, thereby reducing noisy supervision caused by false negatives and enabling the recommender to learn user preferences more reliably.

\definecolor{myboxcolor2}{RGB}{191,128,191}
\newtcolorbox{mybox3}[1]{
	colback=myboxcolor2!5!white,
	colframe=myboxcolor2!75!black,
	fonttitle=\bfseries,
	title=#1,
	halign title=center, 
	halign=left,    
	left=0.5mm,
	right=0.5mm
}
\begin{mybox3}{\small Condensed System Prompt $\mathcal{S}_P$: False Negative Identification}
	\label{prompt: S_P}
	\textit{\textbf{Role:} You are a sample classification module for a recommender system.}
	
	\textit{\textbf{Input:} For each user, you receive the user's historically interacted items and candidate items.
		Each item is represented only by two hierarchical path encodings: collaborative-structure path encodings and latent semantic path encodings.}
		
	\textit{\textbf{Task:} Your task is to classify each candidate item into either \texttt{"positive"} or \texttt{"negative"} based solely on path encoding similarity, which is determined by the shared prefix length between paths (longer shared prefix means higher similarity).}
		
	\textit{\textbf{Output:} Your output is a JSON array where each element is of the form \{"item\_id":<int>, "label":"positive"|"negative"\}}.
	
\end{mybox3}

As specified in the system prompt, DTL-FNI performs direct LLM-based classification over dual-tree structured encodings without explicit chain-of-thought (CoT) prompting~\cite{wei2022chain}. 
Specifically, the LLM compares the collaborative-structural and latent-semantic path patterns between candidate items and users' historical positives, and directly outputs the predicted labels.
This design is mainly motivated by efficiency: CoT prompting generates substantial intermediate reasoning tokens for each user-candidate classification, greatly increasing offline inference time. 
For example, under the same setting on Amazon-sports, CoT prompting is estimated to take 324,247 seconds based on elapsed time and completed inference progress, whereas  direct-classification prompt takes only 1,944 seconds.
This approximately 167$\times$ inference overhead makes CoT prompting unsuitable for DTL-FNI.
Given that direct classification already achieves strong false-negative identification performance, we adopt it as a trade-off between effectiveness and efficiency.

\subsection{Dual-Tree Guided Multi-View Hard Negative Sampling (DT-MHNS)}

We now introduce DT-MHNS, a dual-tree guided multi-view hard negative sampling module that combines user–item preference scores with item–item structural and semantic similarities to mine harder negatives, thereby enhancing the model’s discriminative capability.
For each user $u$ and its positive item $i^+$, we first randomly sample $n$ unobserved items to construct a candidate set $\varepsilon _u$. For each candidate item $i \in \varepsilon_u$, we follow the state-of-the-art method MixGCF~\cite{MixGCF} and use mixup~\cite{mixup} to generate a more challenging hard negative embedding:
\begin{equation}
{\tilde e_{{i}}} = \lambda {e_{i^+}} + (1 - \lambda ){e_{{i}}},\;\;\lambda  \sim U\left( {0,1} \right).
\end{equation}
We compute the preference score between the user embedding $\mathbf{e}_u$ and the hard negative embedding $\tilde{\mathbf{e}}_{i}$ using the inner product: ${\hat r_{u,i}} = e_u^T{\tilde e_i}.$
Considering that items may have different path encoding depths, we measure the similarity between items $i$ and $j$ using the normalized longest common prefix (LCP) length~\cite{karkkainen2017string}, defined as:
\begin{equation}
	\mathrm{sim}(i, j) = 
	\frac{
		\max\{\, l \mid \pi(i)_{1:l} = \pi(j)_{1:l} \,\}
	}{
		\min\bigl(|\pi(i)|,\,|\pi(j)|\bigr)
	}
\end{equation}
where $|\pi(i)|$ denotes the effective length of the path encoding $\pi(i)$. Based on this definition, we compute collaborative-structural and latent-semantic path similarities between the candidate item $i$ and the positive $i^+$, denoted by $\mathrm{sim}^{c}(i^+, i)$ and $\mathrm{sim}^{s}(i^+, i)$, respectively.
 
Next, we unify the user–item preference signal and item–item dual-tree path encoding similarities into a scoring function for hard negative selection.
Since the hardness of a negative sample is fundamentally user-dependent, we use the user–item preference score as the primary hardness signal.
In contrast, the dual-tree path similarities $\mathrm{sim}^{c}(i^+, i)$ and $\mathrm{sim}^{s}(i^+, i)$ characterize the item–item proximity between the candidate negative and the positive item from collaborative-structural and semantic perspectives, providing item–item priors that complement the preference signal. Therefore, we introduce weights $\alpha_c$ and $\alpha_s$ to fuse these signals and control their contributions when measuring the overall hardness of a candidate negative.
\begin{equation}
Score(u, i^+, i) =
\mathrm{Norm}_{\varepsilon_u}(\hat r_{u,i})
+ \alpha_c \cdot \mathrm{sim}^{c}(i^+, i)
+ \alpha_s \cdot \mathrm{sim}^{s}(i^+, i)
\end{equation}
where $\mathrm{Norm}_{\varepsilon_u}(\hat r_{u,i})$ denotes min–max normalization~\cite{henderi2021comparison} applied to all preference scores $\{\hat r_{u,j} \mid j \in \varepsilon_u\}$, rescaling them to $[0,1]$ to ensure comparability with the path encoding similarities and avoid scale-induced fusion bias.

Finally, based on $\mathrm{Score}(u, i^+, i)$, we redefine the negative sampling distribution $q(\cdot \mid u, i^+)$ and, following prior work~\cite{ANS, DNS, MixGCF}, we approximate sampling from this distribution by selecting the hardest negative in the candidate set:
\begin{equation}
	i^- = \arg\max_{i \in \varepsilon_u} Score(u, i^+, i).
\end{equation}
We take the mixup embedding $\tilde{e}_{i^{-}}$ of the selected negative item $i^{-}$ (constructed by Eq.~(10)) as the final hard negative embedding and use it in the subsequent BPR optimization.

It is important to emphasize that the dual-tree path-encoding similarities are item-index–level proximity priors that, together with the preference score, determine which candidate item is selected as the negative sample $i^-$. Although BPR optimization uses the mixup embedding $\tilde{e}_{i^-}$ of the selected negative, this embedding is still derived from, and uniquely associated with, the chosen negative item $i^-$. Therefore, the dual-tree similarities guide the selection of $i^-$ such that the chosen negative tends to be closer to the positive item in both collaborative-structural and latent-semantic spaces. On top of this, the mixup embedding makes the prediction of this negative item for the target user more challenging, thereby enhancing the model’s discriminative capability. The effectiveness of incorporating dual-tree similarities into hard negative sampling is further validated in our ablation study (Section~4.4).

Algorithm~\ref{alg:dtlns_training} describes the overall pipeline of DTL-NS. (Lines~1–9) use the LLM to identify false negatives and convert them into positives to augment the training set for subsequent optimization. (Lines~10–22) then train the recommender with DTL-NS, where (Lines~14–18) perform dual-tree guided multi-view hard negative sampling to select informative negatives.

\begin{algorithm}[t]
	\renewcommand{\algorithmicrequire}{\textbf{Input:}}
	\caption{DTL-NS Training Process}
	\label{alg:dtlns_training}
	\begin{algorithmic}[1]
		\REQUIRE Training set $\mathcal{D}_{\text{train}}\subseteq \mathcal{U}\times\mathcal{I}$; Pretrained recommender $g$; LLM classifier $f_{\mathrm{LLM}}$ with system prompt $S_p$; Collaborative-view weight $\alpha_c$; Semantic-view weight $\alpha_s$; Implicit CF model $\text{rec}\left(  \cdot  \right)$; Number of negative candidate $n$.
		\STATE Obtain dual-tree path codes $\pi^c(\cdot),\pi^s(\cdot)$ call Algorithm~\ref{alg:offline_dualtree}.
		\STATE Construct candidate item sets $C_{\mathcal U}=\{C_u\mid u\in\mathcal{U}\}$ using $g$.
		\STATE $\hat{\mathcal{D}} \leftarrow \mathcal{D}_{\text{train}}$.
		\FOR{each user $u\in\mathcal{U}$}
		\STATE $I_u^+ \leftarrow \{ i \mid (u,i)\in \mathcal{D}_{\text{train}} \}$.
		\STATE Construct prompts ${\rm P}(u,i)$ based on $I_u^+$ and $C_u$ using (7).
		\STATE Use $f_{\mathrm{LLM}}$ to identify false negatives based on ${\rm P}(u,i)$ and collect LLM-verified positives $\hat P_u$ using (9).
		\STATE $\hat{\mathcal{D}} \leftarrow \hat{\mathcal{D}} \cup \{(u,i)\mid i\in \hat P_u\}$.
		\ENDFOR
		\FOR{$t = 1, 2,…,T$}
		\STATE Sample a mini-batch of pairs $D_{batch}$ from $\hat{\mathcal{D}}$.
		\STATE Initialize loss $\ell  = 0$.
		\FOR{each $\left( {u,i^+} \right) \in D_{batch}$}
		\STATE Uniformly sample $n$ negatives to form candidate set ${\varepsilon _u}$.
		\STATE Obtain mixup negative embeddings $\{\tilde{\mathbf e}_{i}\}_{i\in\varepsilon_u}$ by (10).
		\STATE Look up $\pi^c(i^+),\pi^s(i^+)$ and $\{(\pi^c(i),\pi^s(i))\}_{i\in\varepsilon_u}$ from dual-tree path codes.
		\STATE Compute $Score(u,i^+,i)$ for each $i\in\varepsilon_u$ using (12).
		\STATE $i^{-} \leftarrow \arg\max_{i \in \varepsilon_u} Score(u,i^+,i)$.
		
		\STATE $\ell = \ell - \log \sigma \left( e_{i^+} \cdot e_u - \tilde e_{i^-} \cdot e_u \right)$.
		\ENDFOR
		\STATE Update $\theta$ of $\text{rec}\left(  \cdot  \right)$ by descending the gradients ${\nabla _\theta }\ell $.
		\ENDFOR
	\end{algorithmic}
\end{algorithm}

\subsection{Time Complexity}

The time complexity of DTL-NS mainly comes from two modules: DTL-FNI and DT-MHNS. DTL-FNI involves two offline operations whose costs are difficult to
characterize by a standard time-complexity term: pretraining a
recommender $g$ for latent semantic embeddings,
and performing LLM inference for false-negative identification.
We denote these costs by $\mathcal C_{\mathrm{pre}}(g)$ and
$\mathcal C_{\mathrm{LLM}}$, respectively, and report their wall-clock runtime in Section~4.2.
Apart from these two operations, constructing the sparse item-item similarity matrix $\mathbf W^c$ and computing collaborative-structure item embeddings cost
$O(\sum_{u\in\mathcal U}|\mathcal I_u^+|^2+\mathrm{nnz}(\mathbf W^c)d_c)$,
and constructing $k$-ary index trees costs
$O(|\mathcal I|(d_cL_c+d_sL_s))$, where $d_c,d_s$ are embedding dimensions, $L_c,L_s$ are tree depths, and $\mathrm{nnz}(\mathbf W^{c})$ denotes the number of non-zero entries in $\mathbf W^{c}$.
DT-MHNS mainly includes constructing the candidate negative sets, with a cost of $O(nT)$,
and computing the overall hardness scores, with a cost of $O(n(d_e+L_c+L_s)  T)$,
where $n$ denotes the size of the candidate negative set, $d_e$ denotes the item embedding dimension during training, and $T$ denotes the total number of processed training instances.
Therefore, complexity of DT-MHNS is approximately
$O(n(d_e+L_c+L_s)T)$.
The overall time complexity of DTL-NS is 
$O(\mathcal C_{\mathrm{pre}}(g)
+\mathcal C_{\mathrm{LLM}}
+\sum_{u\in\mathcal U}|\mathcal I_u^+|^2
+\mathrm{nnz}(\mathbf W^{c})d_c
+|\mathcal I|(d_cL_c+d_sL_s)
+Tn(d_e+L_c+L_s)).$

\section{Experiments}

To demonstrate the effectiveness of our proposed DTL-NS, we conducted extensive experiments and answered the following questions: 
\textbf{(RQ1)} How does DTL-NS perform in comparison to existing negative sampling and LLM-based recommendation methods? \textbf{(RQ2)} How do the key hyperparameters of DTL-NS affect recommendation performance? \textbf{(RQ3)} How do the key components and design choices of DTL-NS contribute to its overall performance?
\textbf{(RQ4)} Does DTL-NS remain effective as a plug-and-play framework across different backbones, negative sampling methods, and LLMs?

\subsection{Experimental Setup}

\begin{table}[t]
	\centering
	
	\vspace{-0.1cm}
	
	\caption{Dataset Statistics}

\vspace{-0.2cm}
	
	\begin{tabular}{ccccc}
		\toprule
		Dataset & Users & Items & Interactions & Density  \\
		\midrule
		Amazon-sports   & 15,138  & 8,439  & 210,439  & 0.165\%\\
		Amazon-toys    & 15,306   & 10,074   & 247,181  & 0.160\%  \\
		Yelp& 20,206  & 15,674  & 468,033 & 0.148\%  \\
		\bottomrule
	\end{tabular}
	
\vspace{-0.2cm}
	
\end{table}

\newcommand{\std}[1]{{\scriptsize$\pm$#1}}

\begin{table*}[t]
	\centering
	
	\caption{Performance comparison on three real-world datasets (\%). The best results are \textbf{bold}. Results are reported as mean $\pm$ standard deviation over five runs.}
	
	\vspace{-0.2cm}
	
	\resizebox{\linewidth}{33.5mm} {
		\setlength{\tabcolsep}{0.7mm}{
			\begin{tabular}{c|cccc|cccc|cccc}
				\toprule
				\multirow{1}{*}{Dataset} 
				& \multicolumn{4}{c|}{Amazon-sports} 
				& \multicolumn{4}{c|}{Amazon-toys} 
				& \multicolumn{4}{c}{Yelp} \\
				\cmidrule(r){1-1} 
				\cmidrule(r){2-5} 
				\cmidrule(r){6-9} 
				\cmidrule(r){10-13}
				\multirow{1}{*}{Method} 
				& R@10 & N@10 & R@20 & N@20
				& R@10 & N@10 & R@20 & N@20
				& R@10 & N@10 & R@20 & N@20 \\
				\midrule
				RNS        
				& 8.74\std{0.13} & 5.94\std{0.05} & 12.19\std{0.09} & 6.95\std{0.04}
				& 7.86\std{0.15} & 5.07\std{0.08} & 11.48\std{0.16} & 6.13\std{0.03}
				& 8.08\std{0.08} & 5.25\std{0.09} & 13.27\std{0.09} & 6.88\std{0.08} \\
				
				DNS        
				& 10.29\std{0.09} & 7.18\std{0.15} & 13.64\std{0.11} & 8.16\std{0.11}
				& 9.06\std{0.18} & 6.02\std{0.12} & 12.70\std{0.28} & 7.10\std{0.15}
				& 8.57\std{0.17} & 5.57\std{0.06} & 13.76\std{0.07} & 7.20\std{0.04} \\
				
				SRNS       
				& 9.71\std{0.14} & 6.80\std{0.21} & 13.40\std{0.14} & 7.88\std{0.15}
				& 8.10\std{0.06} & 5.54\std{0.08} & 11.98\std{0.14} & 6.64\std{0.09}
				& 8.54\std{0.07} & 5.50\std{0.03} & 13.75\std{0.07} & 7.15\std{0.02} \\
				
				DNS(M,N)   
				& 10.38\std{0.13} & 7.29\std{0.14} & 13.62\std{0.19} & 8.24\std{0.16}
				& 9.26\std{0.01} & 6.30\std{0.10} & 12.49\std{0.28} & 7.27\std{0.10}
				& 8.82\std{0.10} & 5.69\std{0.04} & 14.03\std{0.12} & 7.33\std{0.05} \\
				
				MixGCF     
				& 10.53\std{0.22} & 7.40\std{0.10} & 14.29\std{0.23} & 8.51\std{0.09}
				& 9.60\std{0.16} & 6.33\std{0.07} & 13.24\std{0.23} & 7.41\std{0.09}
				& 8.74\std{0.07} & 5.68\std{0.06} & 14.36\std{0.13} & 7.45\std{0.04} \\
				
				BNS        
				& 9.97\std{0.15} & 6.89\std{0.08} & 13.83\std{0.06} & 8.00\std{0.05}
				& 9.14\std{0.11} & 5.91\std{0.03} & 12.65\std{0.06} & 6.96\std{0.02}
				& 8.62\std{0.15} & 5.56\std{0.08} & 13.72\std{0.13} & 7.20\std{0.10} \\
				
				AHNS       
				& 10.11\std{0.03} & 7.16\std{0.10} & 13.44\std{0.17} & 8.15\std{0.07}
				& 8.84\std{0.13} & 5.85\std{0.06} & 12.47\std{0.11} & 6.94\std{0.03}
				& 8.54\std{0.09} & 5.55\std{0.04} & 13.87\std{0.12} & 7.22\std{0.05} \\
				
				\cmidrule(r){1-13}
				
				KAR        
				& 10.32\std{0.27} & 7.12\std{0.21} & 13.65\std{0.30} & 8.21\std{0.10}
				& 9.14\std{0.25} & 6.27\std{0.18} & 12.53\std{0.27} & 7.19\std{0.13}
				& 8.62\std{0.26} & 5.59\std{0.12} & 14.21\std{0.20} & 7.42\std{0.11} \\
				
				LLMRec     
				& 10.36\std{0.24} & 7.14\std{0.18} & 13.48\std{0.26} & 8.16\std{0.07}
				& 9.17\std{0.12} & 6.31\std{0.16} & 12.46\std{0.23} & 7.15\std{0.12}
				& 8.61\std{0.22} & 5.62\std{0.10} & 13.96\std{0.17} & 7.39\std{0.09} \\
				
				RLMRec     
				& 10.51\std{0.17} & 7.38\std{0.13} & 14.32\std{0.19} & 8.68\std{0.12}
				& 9.55\std{0.15} & 6.46\std{0.11} & 13.17\std{0.15} & 7.46\std{0.08}
				& 8.97\std{0.18} & 5.82\std{0.07} & 14.42\std{0.11} & 7.63\std{0.06} \\
				
				\cmidrule(r){1-13}
				
				HNLMRec    
				& 10.74\std{0.19} & 7.78\std{0.15} & 14.38\std{0.21} & 8.84\std{0.14}
				& 10.33\std{0.17} & 7.70\std{0.13} & 13.79\std{0.17} & 8.73\std{0.09}
				& 9.31\std{0.17} & 6.09\std{0.08} & 14.93\std{0.13} & 7.85\std{0.07} \\
				\cmidrule(r){1-13}
				
				DTL-NS     
				& \textbf{12.46}\std{0.14} & \textbf{9.53}\std{0.10} & \textbf{15.91}\std{0.16} & \textbf{10.53}\std{0.09}
				& \textbf{11.56}\std{0.12} & \textbf{8.98}\std{0.08} & \textbf{14.75}\std{0.12} & \textbf{9.91}\std{0.05}
				& \textbf{10.67}\std{0.12} & \textbf{7.97}\std{0.04} & \textbf{15.66}\std{0.08} & \textbf{9.53}\std{0.03} \\
				
				\bottomrule
			\end{tabular}
	}}

	\vspace{-0.1cm}

\end{table*}

\begin{table}[t]
	\centering

	\caption{Performance comparison on synthetic datasets (\%).}
	
		\vspace{-0.1cm}
	\resizebox{\linewidth}{26mm} {
		\setlength{\tabcolsep}{1.2mm}{
			\begin{tabular}{c|ccc|ccc}
				\toprule
				\multirow{1}{*}{Dataset} 
				& \multicolumn{3}{c|}{Amazon-sports-20\%} 
				& \multicolumn{3}{c}{Amazon-toys-20\%} \\ 
				\cmidrule(r){1-1}
				\cmidrule(r){2-4} 
				\cmidrule(r){5-7}
				\multirow{1}{*}{Method}
				& R@20 & N@10 &N@20 
				& R@20 & N@10 &N@20 \\
				\midrule
				RNS        & 10.81 &5.11 & 6.08 & 10.35 & 4.56 & 5.53 \\
				DNS        & 12.51 & 6.46& 7.41 & 11.26 &5.40 & 6.35 \\
				SRNS     & 12.68 & 5.95& 7.07 & 10.79 &4.90 & 5.86 \\
				DNS(M,N)   & 11.76 & 6.37& 7.16 & 11.57 & 5.65& 6.61 \\
				MixGCF       & 12.70 & 6.46& 7.46 & 11.69 & 5.55& 6.56 \\
				BNS       & 12.75 & 6.38& 7.38 & 11.64 & 5.48 & 6.27 \\
				
				AHNS       & 12.56 & 6.52& 7.48 & 11.33 & 5.39& 6.36 \\
				HNLMRec    & 13.09 & 7.08& 8.11 & 11.95 & 6.16& 7.07 \\
				DTL-NS    & \textbf{14.08} &\textbf{8.74} & \textbf{9.58} & \textbf{13.28} & \textbf{8.05}& \textbf{8.96} \\

				\bottomrule
			\end{tabular}
	}}
	
	\vspace{-0.3cm}
	
\end{table}

\subsubsection{Datasets}
We selected three widely used real-world datasets (Amazon-sports, Amazon-toys, and Yelp) and two synthetic noisy datasets (Amazon-sports-$r$ and Amazon-toys-$r$) to evaluate the effectiveness of DTL-NS. For the noisy datasets, we randomly removed $r\%$ of the training interactions to simulate false negatives.
The Amazon-Sports and Amazon-Toys datasets\footnote{\url{https://cseweb.ucsd.edu/~jmcauley/datasets/amazon_v2/}} contain data spanning from 2014 to 2018 from the Amazon review corpus, corresponding to the Sports\&Outdoors and Toys\&Games categories, respectively.
The Yelp dataset\footnote{\url{https://business.yelp.com/data/resources/open-dataset/}} contains user interaction records from the Yelp platform spanning from 2015 to 2018.
Following prior work~\cite{NGCF, lightgcn, MCNS}, we adopt the 10-core setting for all datasets to ensure that each user and item has at least ten interactions, and then randomly split the data into training, validation, and test sets with a ratio of 8:1:1. 
The dataset statistics are shown in Table 1. We use the widely adopted metrics, Recall@k and NDCG@k to evaluate model performance, with k values set to \{10, 20\}.

\subsubsection{Baselines}
We compare DTL-NS with eight negative sampling methods: the commonly used RNS~\cite{BPR}, three hard negative sampling methods (DNS~\cite{DNS}, MixGCF~\cite{MixGCF}, DNS(M,N)~\cite{DNS(MN)}), three methods targeting false negatives (SRNS~\cite{SRNS}, BNS~\cite{liu2023bayesian}, AHNS~\cite{AHNS}) and a LLM-based negative sampling method HNLMRec~\cite{zhao2025HNLMRec}.
In addition, we include three representative LLM-enhanced recommenders (KAR~\cite{xi2024kar}, LLMRec~\cite{wei2024llmrec}, and RLMRec~\cite{ren2024rlmrec}), as complementary baselines. 
Note that our main comparisons focus on negative sampling methods, since DTL-NS is a negative sampling strategy; LLM-enhanced recommenders follow a different paradigm and are included only as a reference.
For fair comparison, unless otherwise specified, we use the hyperparameter settings reported in the original paper or released code when a baseline is evaluated on the same dataset as in the original study; for other datasets, we tune its key hyperparameters on the validation set.

\begin{figure}[t]
	\centering
	\includegraphics[width=\linewidth]{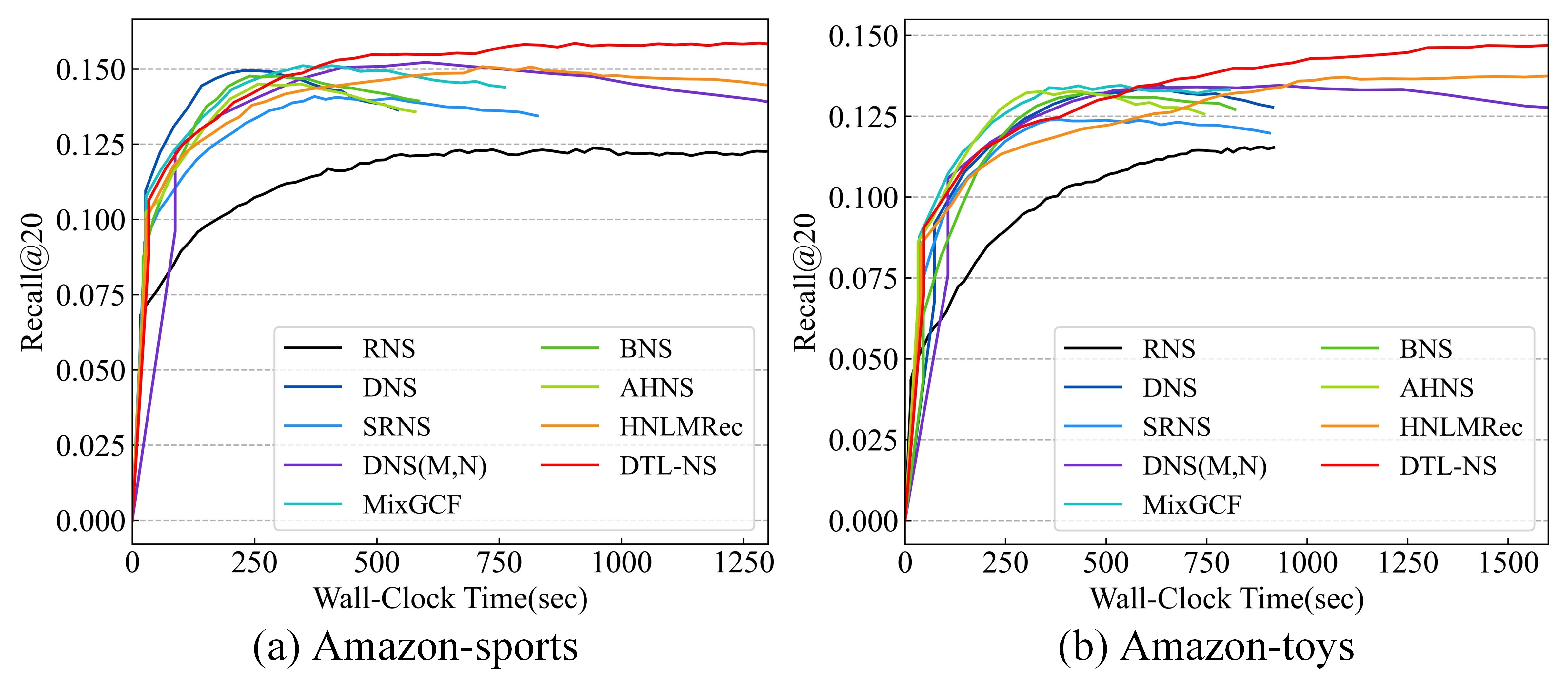}
	
		\vspace{-0.2cm}
	
	\caption{Recall@20 vs. wall-clock time (in seconds).}
	
	\vspace{-0.4cm}
	
\end{figure}

\subsubsection{Implementation Details}
Unless otherwise specified, we use LightGCN~\cite{lightgcn} as the default backbone. 
Llama-3.1-8B-Instruct is adopted to perform inference and identify false negatives. We initialize 64-dimensional embeddings with Xavier initialization~\cite{Xavier} and optimize the model using Adam~\cite{adam} with a learning rate of 0.001. The mini-batch size is 2048, the L2 regularization coefficient is 0.0001, and the number of layers is set to 3. 
In DTL-FNI, dual hierarchical index trees are constructed with a branching factor of 4 and a leaf-size threshold of 30. We set the spectral embedding dimension to 32 for the collaborative-structure tree.
We also use LightGCN as the pretrained recommender to build the candidate set for LLM classification, and set the candidate size $|\mathcal{C}_u|$ to 20. To avoid data leakage, all held-out validation and test interactions are excluded from the candidate pool before LLM classification, since these interactions may otherwise be ranked as high-scoring candidates by the pretrained recommender. 
The numbers of LLM-identified false negatives that are converted into new positives are 193,550 (Amazon-sports), 200,815 (Amazon-toys), and 145,097 (Yelp), respectively.
For fair comparison, all LLM-based baselines in our experiments use Llama-3.1-8B-Instruct. For methods adopting a two-stage negative sampling strategy, we set the candidate negative pool size $|\varepsilon_u|$ to 10.

\subsubsection{Experiment environment.}
Our experiments are conducted on a Linux server with Intel(R) Xeon(R) Gold 6240 CPU@2.60GHz, 251G RAM and 1 NVIDIA A100 80GB PCIe. Our proposed DTL-NS is implemented in Python 3.10.13, torch 2.9.0, scikit-learn 1.6.1, numpy 1.26.4, scipy 1.10.0, transformers 4.57.3, vllm 0.12.0, and CUDA 12.4.0. Our source code is available in the anonymous repository https://anonymous.4open.science/r/DTL-NS-7EF1/.

\subsection{Performance Comparison (RQ1)}

Table~2 reports the performance of DTL-NS and the baselines on real-world datasets. 
The results indicate that DTL-NS achieves significant improvements over strong baselines.
Such substantial gains mainly stem from its ability to identify and convert false negatives into reliable positive supervision.
Specifically, DTL-NS encodes collaborative-structural and latent semantic signals with hierarchical index-tree representations, enabling the LLM to perform classification over structured item encodings and effectively identify false negatives among unobserved items.
The identified false negatives are then converted into additional positives to augment training, effectively transforming noisy signals into reliable supervision.
In contrast, negative sampling baselines either ignore false negatives or merely avoid sampling them, leaving their potential as positive supervision underutilized.
For LLM-enhanced baselines, the improvements largely rely on textual signals, which can be noisy and may not consistently reflect users' true preferences.
Moreover, DTL-NS jointly considers item-item structural/semantic similarities and user-item preference scores to select more challenging hard negatives, further enhancing the model's discriminative ability.
Together, these designs provide cleaner and more informative training signals, leading to substantial performance gains.

Table~3 reports the performance of DTL-NS and the negative sampling baselines on the synthetic noisy datasets. We can see that DTL-NS consistently achieves the best performance, with even more pronounced improvements than those observed on the real-world datasets. For example, on the Amazon-toys, and corresponding synthetic dataset, Recall@20 and NDCG@20 improvements over the strongest baseline increase from 6.96\% to 11.12\% and from 13.52\% to 26.73\%, respectively. These results further demonstrate that DTL-NS can effectively identify false negatives.

Beyond effectiveness, we analyze the efficiency of DTL-NS, which consists of an offline DTL-FNI stage and a training-time DT-MHNS stage.
We first measure the runtime of DTL-FNI for identifying false negatives across all datasets. 
Using vLLM~\cite{vLLM}, which supports optimized KV-cache management and efficient batch scheduling, DTL-FNI takes 3,644, 3,128, and 6,406 seconds in total on the Amazon-sports, Amazon-toys, and Yelp datasets, respectively. 
The dominant costs come from LLM inference and recommender pretraining for candidate construction. 
The former takes 1,937, 2,186, and 3,641 seconds on the three datasets, respectively, while the latter takes 1,699, 931, and 2,749 seconds.
In contrast, the cost of iterative $k$-means-based construction of the $k$-ary hierarchical index trees and item encoding is negligible, taking only 8, 11, and 16 seconds, respectively. 
This is because the hierarchical index trees are constructed at the item level, and the branching factor and leaf-size threshold keep the tree depth and the number of clustering operations limited.
Since DTL-FNI is a one-time offline process, this cost is practical in our setting.
We then compare the average per-epoch training time with and without the multi-view hard negative selection in DT-MHNS on Amazon-sports, Amazon-toys, and Yelp.
The average per-epoch forward-pass time increases slightly from 1.26s, 1.51s, and 1.93s to 1.57s, 1.73s, and 2.47s, indicating that the additional overhead of multi-view negative selection is marginal.
Finally, we compare the overall training efficiency of DTL-NS with negative sampling baselines. 
As shown in Figure~3, DTL-NS is not the fastest method to reach its best performance. This is mainly because DTL-NS is the only method among them that explicitly uses potential false negatives as positive supervision, which enlarges the positive training set and thus leads to relatively longer training time. 
Nevertheless, this design enables the model to learn user preferences more accurately, resulting in more stable convergence and better final performance. Therefore, this trade-off between training efficiency and recommendation accuracy is acceptable, and further demonstrates the practical value of DTL-NS.

\begin{figure}[t]
	\centering	
	\includegraphics[width=\linewidth]{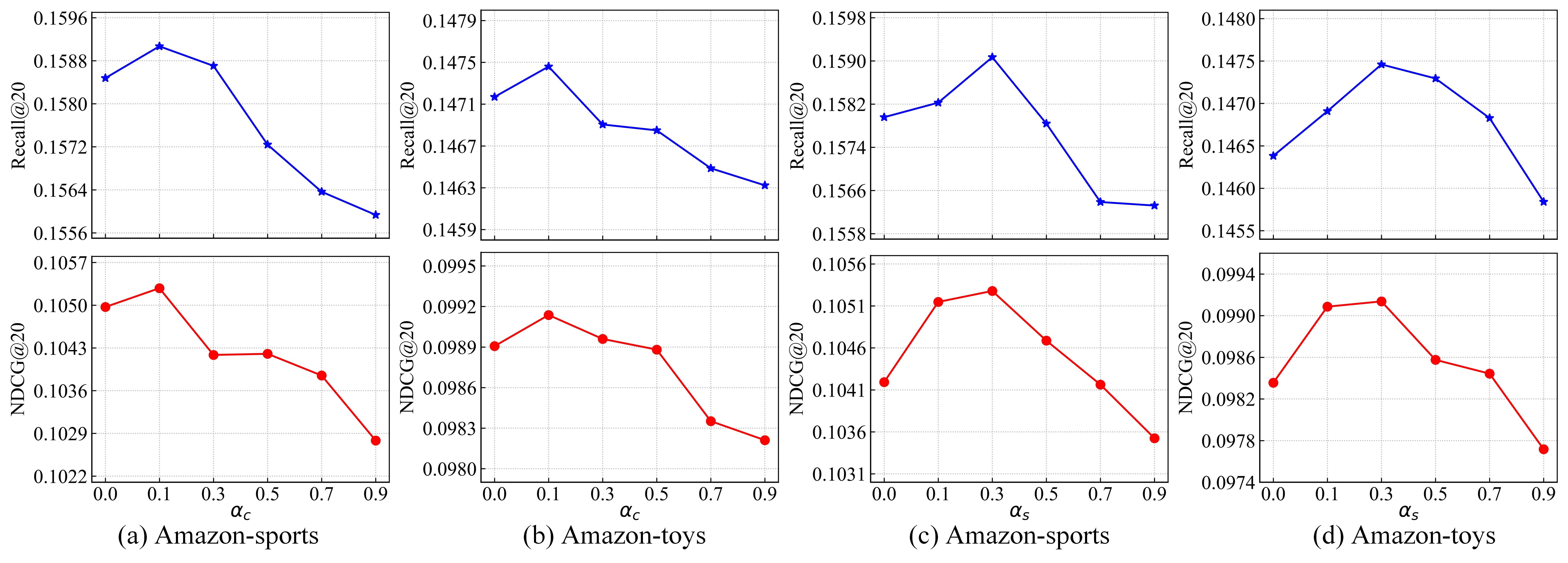}
	
	\vspace{-0.2cm}
	
	\caption{The impact of $\alpha_c$ and $\alpha_s$.}
	
		\vspace{-0.4cm}
\end{figure}

\subsection{Parameter Analysis (RQ2)}

In this section, we analyze the impact of the parameters $\alpha_c$ and $\alpha_s$ in DTL-NS on model performance, where $\alpha_c$ and $\alpha_s$ control the contributions of the collaborative-structural and latent-semantic signals to the overall hardness score, respectively.
We vary each coefficient in \{0.0, 0.1, 0.3, 0.5, 0.7, 0.9\}.

Figure~4 illustrates the impact of $\alpha_c$ and $\alpha_s$. We observe a similar trend for both parameters: as the weight increases, the performance first improves and then degrades, with better results achieved when the weight is set to a relatively small value.
In particular, setting $\alpha_c$ to 0.1 and $\alpha_s$ to 0.3 yields the highest Recall@20 and NDCG@20 on both Amazon-sports and Amazon-toys datasets.
This is because moderately incorporating the item-item priors (i.e., $\mathrm{sim}^{(c)}(i^+, i)$ for collaborative-structural signals and $\mathrm{sim}^{(s)}(i^+, i)$ for latent-semantic signals) helps select hard negatives that are both highly confusing for the target user (with high preference scores) and close to the corresponding positive item, thereby providing more informative learning signals and inducing more discriminative parameter updates, which ultimately improves model performance.
Note that the hardness of a negative sample is fundamentally user-dependent, and the user-item preference score should therefore remain the primary hardness signal.
When $\alpha_c$ or $\alpha_s$ is large, the corresponding item-item similarity term receives a relatively high weight in the overall hardness score, which can weaken the preference signal and may bias the sampler toward negatives that are close to the positive item but not sufficiently challenging for the target user. As a result, the selected hard negatives become less discriminative, leading to performance degradation.

\subsection{Ablation Study (RQ3)}

%
%
%
%

\begin{table}[t]
	\centering
	\caption{Ablation results of DTL-NS and its variants.}
	\vspace{-0.1cm}
	\label{tab:ablation}
	\setlength{\tabcolsep}{3.5pt}
	\begin{tabular}{c|ccc|ccc}
		\toprule
		\multirow{2}{*}{Method}
		& \multicolumn{3}{c|}{Amazon-sports} 
		& \multicolumn{3}{c}{Amazon-toys} \\
		\cmidrule(lr){2-4}\cmidrule(lr){5-7}
		& R@20 & N@10 & N@20 & R@20 & N@10 & N@20 \\
		\midrule
		LightGCN & 12.19 & 5.94 & 6.95 & 11.48 & 5.07 & 6.13 \\ 
		\midrule
		DTL-NS$_{\mathbf{C}}$ & 15.57 & 9.13 & 10.17 & 14.25 & 8.32 & 9.33 \\
		DTL-NS$_{\mathbf{S}}$ & 15.48 & 9.09 & 10.13 & 14.17 & 8.27 & 9.27 \\ 
		\midrule
		DTL-FNI & 15.59 & 9.00 & 10.08 & 13.96 & 8.47 & 9.39 \\
		DT-MHNS & 14.56 & 7.53 & 8.66 & 13.41 & 6.52 & 7.59 \\ 
		\midrule
		RuleFNI & 14.31 & 7.55 & 8.63 & 12.89 & 7.32 & 8.25 \\
		KNNFNI & 14.15 & 7.42 & 8.48 & 13.36 & 7.59 & 8.54 \\ 
		XGBFNI & 14.65 & 8.22 & 9.28 & 13.61 & 7.84 & 8.78 \\
		\midrule
		DTL-NS & \textbf{15.91} & \textbf{9.53} & \textbf{10.53} & \textbf{14.75} & \textbf{8.98} & \textbf{9.91} \\
		\bottomrule
	\end{tabular}
	\vspace{-0.3cm}
\end{table}

In this section, we conduct experiments on the Amazon-sports and Amazon-toys datasets and design the following variant methods to validate the effectiveness of the key components and design choices in DTL-NS.
\textbf{DT-MHNS}: keeps only DT-MHNS for training-time multi-view hard negative sampling.
\textbf{DTL-FNI}: keeps only DTL-FNI to augment the training set, and trains the backbone with random negative sampling.
\textbf{DTL\text{-}NS$_{\mathbf{C}}$}: uses only the collaborative structural tree for false-negative identification and multi-view hard negative sampling.
\textbf{DTL\text{-}NS$_{\mathbf{S}}$}: relies solely on the latent semantic tree for the same process.
To examine whether the gains of DTL-FNI stem from higher-quality LLM-based false-negative identification rather than simply introducing more augmented positives, we further introduce three scale-matched no-LLM variants.
All three variants operate on the same candidate sets constructed by the same pretrained recommender as DTL-FNI, i.e., LightGCN in our default setting; they assign each candidate item a false-negative score and identify higher-scoring candidates as potential false negatives.
\textbf{RuleFNI}: is a deterministic rule-based no-LLM variant. It computes the score using the maximum dual-tree path similarity between the candidate item and the user's historical positives.
\textbf{KNNFNI}: is a non-parametric retrieval-based no-LLM variant. It computes the score based on the weighted overlap between the candidate item's dual-tree item-KNN sets and the user's historical positives.
\textbf{XGBFNI}: serves as a strong supervised no-LLM competitor. 
It extracts non-textual features for each candidate, including collaborative-tree and semantic-tree path-similarity statistics, user-history features, and item-popularity features, and trains an XGBoost~\cite{chen2016xgboost} model to estimate its probability of being a false negative. 
In our experiments, the parameters of these no-LLM variants are adjusted to control the scale of identified potential false negatives relative to DTL-FNI.
Specifically, the numbers of potential false negatives identified by RuleFNI, KNNFNI, and XGBFNI are 196,167, 196,087, and 195,972 on Amazon-sports, respectively, and 204,513, 208,449, and 205,457 on Amazon-toys, respectively.
These numbers are slightly larger than those identified by DTL-FNI, i.e., 193,550 and 200,815 on Amazon-sports and Amazon-toys, respectively.
This scale-matched design ensures that the no-LLM variants introduce no fewer additional positive samples than DTL-FNI. 
Therefore, if DTL-FNI still achieves better recommendation performance, the improvement cannot be attributed to enlarging the positive set, but rather indicates the higher quality of LLM-based false-negative identification.

\begin{table}[t]
	\centering
	\caption{Performance comparison with different backbones.}
	
		\vspace{-0.1cm}
	
	\setlength{\tabcolsep}{3.5pt}
	\begin{tabular}{c c|ccc|ccc}
		\toprule
		
		& Dataset
		& \multicolumn{3}{c|}{Amazon-sports} 
		& \multicolumn{3}{c}{Amazon-toys} \\  
		\cmidrule(lr){2-2}\cmidrule(lr){3-5}\cmidrule(lr){6-8}
		& Method
		& R@20 & N@10 & N@20 
		& R@20 & N@10 & N@20 \\  
		\midrule
		
		\multirow{8}{*}{\rotatebox[origin=c]{90}{MF}}
		& RNS      & 10.20 & 5.10 & 5.92 & 8.59 & 3.85 & 4.59 \\ 
		& DNS      & 10.92 & 5.83 & 6.64 & 10.03 & 4.79 & 5.67 \\ 
		& SRNS     & 10.86 & 5.84 & 6.67 & 9.87 & 4.72 & 5.55 \\ 
		& DNS(M,N) & 10.65 & 6.01 & 6.73 & 10.40 & 5.29 & 6.11 \\ 
		& BNS & 10.91& 5.76 & 6.56 & 9.92 & 4.61 & 5.45 \\ 
		& AHNS     & 11.23 & 5.89 & 6.71 & 10.18 & 4.85 & 5.72 \\ 
		& HNLMRec  & 12.11 & 6.35 & 7.21 & 12.13 & 5.56 & 6.57 \\ 
		& DTL-NS   & \textbf{14.52} & \textbf{8.19} & \textbf{9.16} 
		& \textbf{13.34} & \textbf{7.20} & \textbf{8.22} \\ 
		
		\midrule
		
		\multirow{9}{*}{\rotatebox[origin=c]{90}{SimGCL}}
		& RNS      & 13.05 & 6.30 & 7.42 & 11.91 & 5.62 & 6.65 \\ 
		& DNS      & 14.16 & 7.14 & 8.24 & 12.93 & 6.15 & 7.28 \\ 
		& SRNS     & 13.78 & 6.51 & 7.71 & 12.34 & 5.67 & 6.78 \\ 
		& MixGCF   & 14.80 & 7.40 & 8.59 & 13.17 & 6.32 & 7.43 \\ 
		& DNS(M,N) & 14.33 & 7.45 & 8.56 & 12.92 & 6.36 & 7.40 \\ 
		& BNS &13.69 & 6.70 & 7.85 & 12.52 & 5.90 & 6.95 \\ 
		& AHNS     & 13.66 & 7.01 & 8.09 & 12.82 & 6.14 & 7.22 \\ 
		& HNLMRec  & 14.75 & 7.51 & 8.62 & 13.25 & 6.53 & 7.56 \\ 
		& DTL-NS   & \textbf{16.17} & \textbf{9.66} & \textbf{10.71} & \textbf{14.83} & \textbf{9.06} & \textbf{9.99} \\ 
		
		\bottomrule
	\end{tabular}
	
		\vspace{-0.3cm}
	
\end{table}

Table~4 reports the ablation results of DTL-NS and its variants on Amazon-sports and Amazon-toys, with LightGCN included as the backbone baseline. 
Both DTL-FNI and DT-MHNS consistently improve over LightGCN, validating the effectiveness of DTL-FNI and DT-MHNS; 
moreover, combining them in DTL-NS achieves the best performance, indicating that DTL-FNI and DT-MHNS contribute complementary benefits by improving different aspects of the training signal (positive supervision enhancement vs. hard negative mining), resulting in further gains beyond using either module alone.
Notably, DTL-FNI yields a larger performance gain than DT-MHNS.
This aligns with our intuition that DTL-FNI yields larger gains by converting LLM-identified false negatives into positive samples, thereby augmenting explicit positive supervision (i.e., what the user likes) and strengthening the learning of users' global preference signals; in contrast, DT-MHNS mainly reshapes the negative sampling distribution by selecting harder negatives, which primarily refines local decision boundaries via negative supervision (i.e., what the user may not prefer).
In addition, DTL-NS consistently outperforms DTL\text{-}NS$_{\mathbf{C}}$ and DTL\text{-}NS$_{\mathbf{S}}$, indicating that the collaborative structural and latent semantic views are complementary and should be jointly leveraged.
Finally, the three no-LLM variants, i.e., RuleFNI, KNNFNI, and XGBFNI, generally improve over the backbone, indicating that dual-tree path encoding similarities and their derived non-textual features can help identify a portion of potential false negatives.
Notably, although these no-LLM variants introduce a slightly larger number of additional positives than DTL-FNI, DTL-FNI still substantially outperforms them.
This indicates that the gains of DTL-FNI mainly come from the higher-quality false-negative identification enabled by LLM-based classification, rather than merely expanding the positive set.
This advantage can be attributed to the multi-history, multi-candidate, and dual-tree encoding setting, where explicitly designing a deterministic rule or learning a compact decision function that jointly accounts for these factors is non-trivial and may fail to generalize across different users and candidate patterns, thus missing cross-item and cross-view consistency cues.
In contrast, the LLM can aggregate evidence across multiple historical items and both collaborative-structural and latent-representation views, and better handle ambiguous cases by capturing cross-item and cross-view consistency, yielding more accurate identification and stronger gains.

\subsection{Applicability Analysis (RQ4)}

\begin{figure}[t]
	\centering	
	\includegraphics[width=\linewidth]{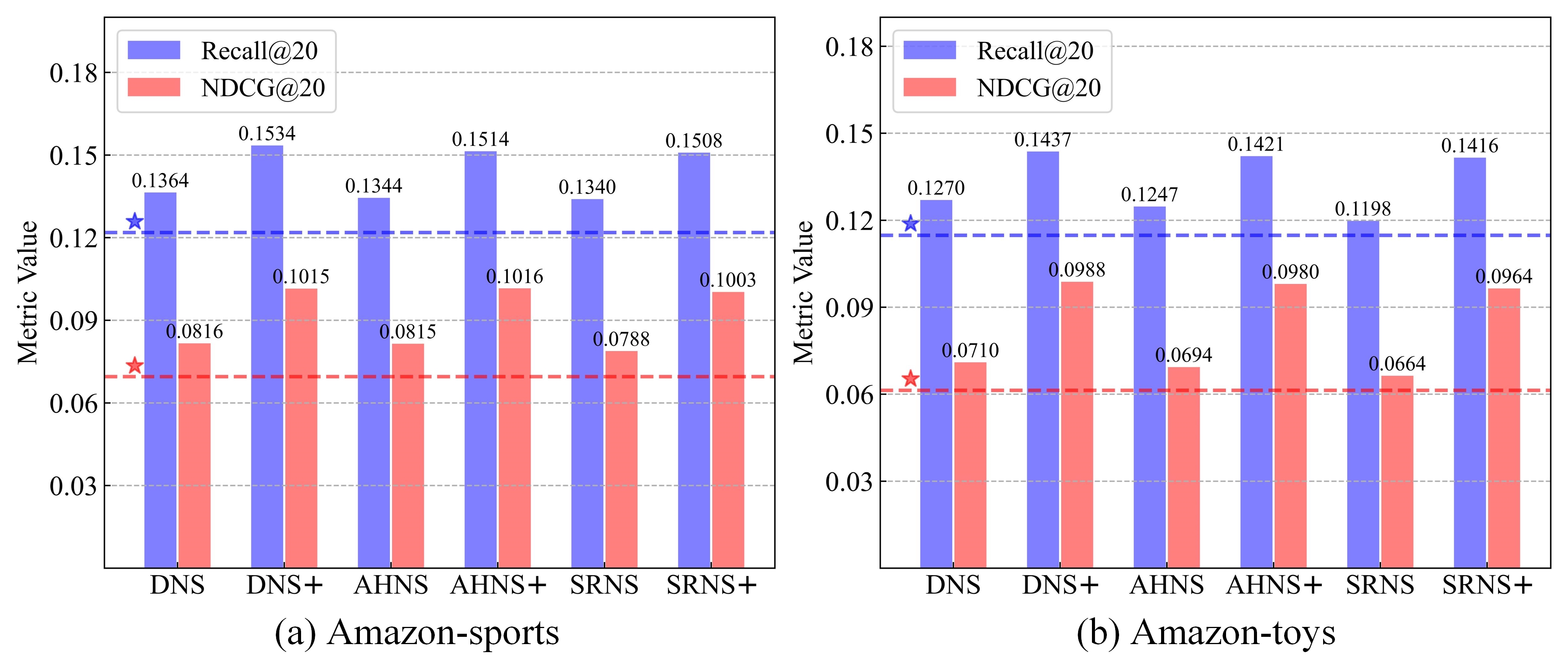}
	
	\vspace{-0.22cm}
	
\caption{Performance comparison under different negative sampling methods.}
	
		\vspace{-0.3cm}
	
\end{figure}

In this section, we investigate the applicability of DTL-NS from three perspectives: whether it can be applied to improve different implicit CF backbones, whether it can enhance existing negative sampling methods, and whether it can maintain effective performance when using different LLMs for false-negative identification.

\textbf{Integration with different implicit CF backbones.}
We first apply DTL-NS to other representative implicit CF backbones, including a widely used matrix factorization (MF)-based CF model~\cite{NCF} and a self-supervised graph CF model~\cite{SimGCL}.
Table~5 reports the performance of DTL-NS and negative sampling baselines on the Amazon-sports and Amazon-toys datasets.
For MF, we omit MixGCF because its hop-mixing component is built upon GNN-based graph propagation, making it not directly applicable to MF-based CF models.
As shown in Table~5, DTL-NS consistently outperforms all baselines under different backbones.
This improvement mainly stems from LLM-assisted false-negative identification, which accurately relabels false negatives as positives, thereby reducing false-negative noise and providing additional reliable positive supervision.
Such explicit augmented supervision is particularly beneficial for relatively simple backbones such as MF, as it helps them better capture users' preferences despite their limited representation capacity, leading to larger performance improvements.
Moreover, we observe that when more expressive backbones are used, most negative sampling methods can further improve their recommendation performance.
This observation is consistent with our intuition: stronger backbones can learn higher-quality user and item representations, which provide more accurate preference estimates and help negative sampling methods select higher-quality negative samples, thereby offering more informative training signals to further improve backbone performance.

\begin{figure}[t]
	\centering	
	\includegraphics[width=\linewidth]{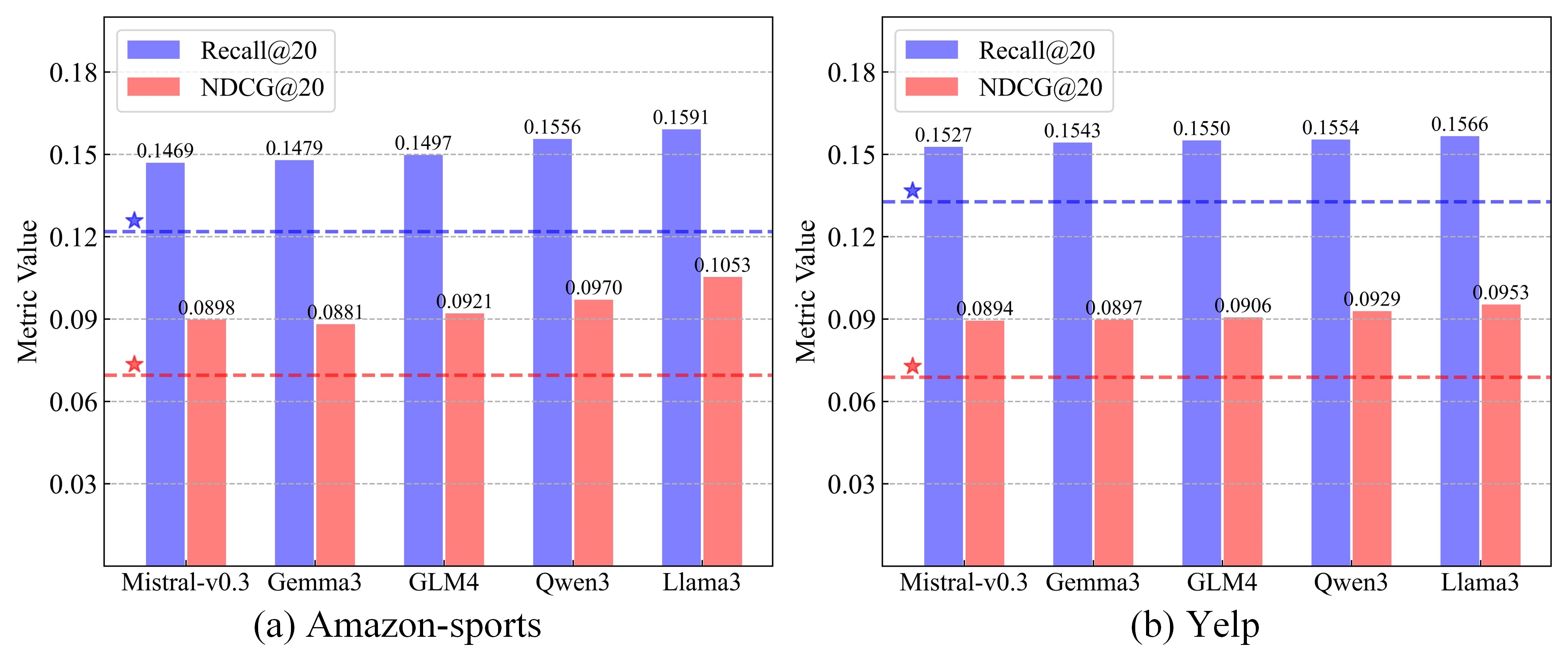}
	
		\vspace{-0.2cm}
	
	\caption{Performance comparison with different LLMs.}
	
		\vspace{-0.3cm}
	
\end{figure}

\textbf{Integration with different negative sampling methods.}
Beyond backbone-level integration, we further integrate DTL-NS into three popular negative sampling methods, DNS, SRNS, and AHNS, yielding DNS${+}$, SRNS${+}$, and AHNS${+}$.
Specifically, we augment the positive set of each sampler with the false negatives identified by DTL-FNI, and incorporate item-item collaborative and semantic similarities from DT-MHNS when computing sampling scores to construct the negative sampling distribution.
As shown in Figure~5, DTL-NS significantly boosts the performance of these methods, mainly because it effectively mitigates their false-negative risk by strengthening positive supervision and simultaneously selects harder negatives.
These results indicate that DTL-NS is not limited to a specific negative sampling strategy, but can serve as a plug-and-play enhancement for existing samplers.

\textbf{Integration with different LLMs.}
Since DTL-FNI relies on LLMs to identify false negatives, we further examine whether DTL-NS can work with different LLMs.
Specifically, we replace Llama-3.1-8B-Instruct with several representative LLMs, including Mistral-7B-Instruct-v0.3~\cite{jiang2023mistral}, Gemma-3-4B-it~\cite{gemma2025gemma3}, GLM-4-9B-Chat-hf~\cite{glm2024chatglm}, and Qwen3-8B~\cite{qwen2025qwen3}, while keeping all other settings unchanged.
As shown in Figure~6, DTL-NS consistently achieves substantial improvements over the backbone across different LLMs, where the blue and red dashed horizontal lines denote the backbone performance in terms of Recall@20 and NDCG@20, respectively.
This indicates that the effectiveness of DTL-NS does not depend on a specific LLM.
Among them, Llama-3.1-8B-Instruct achieves the best overall performance, followed by Qwen3-8B, suggesting that LLMs with stronger instruction-following and semantic discrimination capabilities can identify false negatives more accurately, thereby providing higher-quality positive supervision and bringing larger performance gains.
These results demonstrate that DTL-NS can be flexibly integrated with different LLMs for false-negative identification.  

Overall, these results demonstrate the broad applicability and practical value of DTL-NS across different implicit CF backbones, negative sampling methods, and LLMs.

\section{CONCLUSION}

In this paper, we study LLM-enhanced negative sampling for implicit CF from a new perspective: leveraging LLMs to identify false negatives.
We innovatively introduce hierarchical index trees to encode collaborative-structural and latent-semantic information into structured ID, enabling an LLM to effectively identify false negatives without relying on textual metadata and fine-tuning, thereby improving applicability under limited text and compute.
Then, we convert the identified false negatives into positive supervision, explicitly guiding model to learn user preferences accurately.
Based on item-item hierarchical similarities, we design a multi-view hard negative sampling strategy that mines harder negatives and enhances model's discriminative ability.
Comprehensive experiments demonstrate the effectiveness, generality, and plug-and-play capability of DTL-NS.
This introduces a promising direction for negative sampling: transforming noisy implicit signals into reliable positive supervision via precise false-negative identification, thereby further improving implicit CF recommender performance.

\section{GenAI Usage Disclosure}
The authors used generative AI tools only to assist with language polishing and presentation refinement. 
The authors carefully reviewed all AI-assisted edits and take full responsibility for the content of this manuscript.

\bibliographystyle{ACM-Reference-Format}
\bibliography{citiation}

\end{document}